\documentclass[useAMS,usenatbib,usegraphicx]{mn2e}
\usepackage{graphicx}

\title{Internal shock model for the X-ray flares of Swift J1644+57}
\author[F. Y. Wang and K. S. Cheng] {F. Y. Wang$^{1,2,3}$ and K. S. Cheng$^1$\\
$^1$Department of Physics, The University of Hong Kong,
Pokfulam Road, Hong Kong, China \\
$^2$Department of Astronomy, Nanjing University, Nanjing 210093,
China \\
$^3$Key Laboratory of Modern Astronomy and Astrophysics (Nanjing
University), Ministry of Education, Nanjing 210093, China}
\begin{document}

\maketitle

\begin{abstract}
Swift J1644+57 is an unusual transient event, likely powered by the
tidal disruption of a star by a massive black hole. There are
multiple short timescales X-ray flares were seen over a span of
several days. We propose that these flares could be produced by
internal shocks. In the internal shock model, the forward and
reverse shocks are produced by collisions between relativistic
shells ejected from central engine. The synchrotron emission from
the forward and reverse shocks could dominate at two quite different
energy bands under some conditions, the relativistic reverse shock
dominates the X-ray emission and the Newtonian forward shock
dominates the infrared and optical emission. We show that the
spectral energy distribution of Swift J1644+57 could be explained by
internal shock model.
\end{abstract}


\section{Introduction}
Swift J1644+57 was triggered by the Swift/BAT on 28 March 2011
(Cummings et al. 2011). Swift J1644+57 was initially discovered as a
long-duration gamma-ray burst (GRB 110328A) by the Swift satellite,
but the light curve soon showed that it was quite different. It
remained bright and highly variable for a long period, and
re-trigger the BAT three times over the next 48 hours (Sakamoto et
al. 2011). The isotropic X-ray luminosity of Swift J1644+57 ranges
from $10^{45}-4\times 10^{48}~$erg~s$^{-1}$, and the total isotropic
energy is about $3\times 10^{53}$ erg during the first 30 days after
the BAT trigger (Burrows et al. 2011). From the strong emission
lines of hydrogen and oxygen, Levan et al. (2011) estimate the
redshift of Swift J1644+57 is $z=0.35$. From the astrometric
observation of the X-ray, optical, infrared, and radio transient
with the light-centroid of the host galaxy, it is found that the
position of this source is consistent with arising in the nucleus of
the host galaxy (Bloom et al. 2011; Zauderer et al. 2011).

The X-ray light curve of Swift J1644+57 exhibits repeated extremely
short timescale flares. The flares have rise-times as short as $100$
s (Burrows et al. 2011). These flares are similar as the flares
discovered in the GRB afterglow (Burrows et al. 2005), which may
indicate the same origin of them. The internal shock model can
produce the X-ray flares observed in GRB afterglows (Burrows et al.
2005; Fan \& Wei 2005; Zhang et al. 2006; Yu \& Dai 2009).

After the Swift J1644+57 was discovered, several models were
proposed to explain it, most concentrating on a picture that a main
sequence star was tidally disrupted by passing too close to a
$10^6-10^7M_{\odot}$ black hole (Bloom et al. 2011; Burrows et al.
2011; Cannizzo et al. 2011; Socrates 2011; Shao et al. 2011). Krolik
\& Piran (2011) suggest that this event may be produced by a white
dwarf tidally disrupted by a $10^4M_\odot$ black hole. The process
is as follow: a star is disrupted as it passes near a supermassive
black hole, and much of its mass is distributed into an accretion
disk around the black hole. A powerful jet is then launched. In
these models, the X-ray emission is thought to be produced by
external inverse Compton (EIC) (Bloom et al. 2011) or synchrotron
emission (Burrows et al. 2011). But on the high frequency side, the
Fermi LAT (Campana et al. 2011) and VERITAS upper limits (Aliu et
al. 2011) require that the synchrotron self-Compton (SSC) component
is suppressed by $\gamma-\gamma$ pair production. The soft photons
of $\gamma-\gamma$ pair production are thought to be generated from
the thermal emission of the accretion disk or the disk outflow. In
the SSC model, soft photons originated from the thermal emission of
the accretion disk may not provide an efficient source for the
$\gamma-\gamma$ production. Because the condition of $\gamma-\gamma$
production is $E_XE_\gamma(1-\cos\theta)\geq 2(m_ec^2)^2$, where
$\theta$ is the angle between the directions of soft seed photon and
high-energy photon. Only a fraction of high-energy emission can be
absorbed by soft photons. So the soft photons from the disk outflow
may be a better candidate (Strubbe \& Quataert 2009). In the
synchrotron emission model, the jet must have a strong magnetic
field (Poynting-flux-dominated) and has ongoing in situ acceleration
of electrons (Aliu et al. 2011; Burrows et al. 2011).

In this letter, we use the internal shock model to explain the X-ray
flares of Swift J1644+57. The internal shock produces the prompt
emission of GRB in the standard fireball model (Rees \&
M\'{e}sz\'{a}ros 1994; Paczy\'{n}ski \& Xu 1994). The internal shock
model also the leading model of X-ray flares in GRBs, the external
shock model is very hard to account for the X-ray flares (Burrows et
al. 2005; Fan \& Wei 2005). The central engine of this event may be
formed as follow. When a supermassive black hole tidal disrupts a
star, a disk is formed. The magnetic field could be produced by disk
instability. The disk can then anchor and amplify the seed magnetic
field to a strong ordered poloidal field , which in turn threads the
black hole with debris material in the inner region of the disk. A
large amount of the rotational energy of the black hole can be
extracted via the Blandford-Znajek (BZ) process, which creates a jet
along the rotation axis (Blandford \& Znajek 1977). The magnetic
field lines will break the disk into blobs, so many shells could be
ejected (Cheng \& Lu 2001). When fast shell catches up with early
slow shell, internal shock is generated. Other models of central
engine also discussed, such as the episodic accretion onto a central
object due to a chopped accretion disk (Perna et al. 2006), or
episodic accretion due to a modulation of the accretion flow by a
magnetic barrier (Proga \& Zhang 2006).

The structure of this letter is as follow. In next section, we
describe the dynamics of internal shock arising from a collision
between two shells and the synchrotron radiation of the shocked
electrons. In Section 3, we apply the model to the Swift J1644+57.
Finally, a summary is given in Section 4.

\section{The internal shock model}
\begin{figure}
\begin{center}
\includegraphics[width=0.5\textwidth]{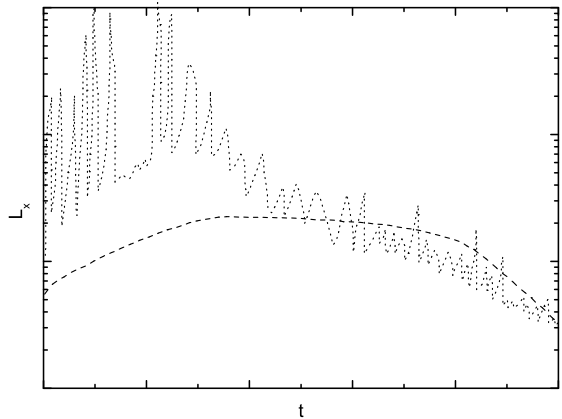} \caption{
Schematic illustration of the model. The dotted line represents the
intensity produced by internal shocks. The dashed line represents
the intensity produced by the external shock. In this figure we
argue that large short-term fluctuations can still occur due to
collisions of few later ejecta when the jet still exists.}
\label{LC}
\end{center}
\end{figure}

The internal shock model has been extensively discussed in
literature (Rees \& M\'{e}sz\'{a}ros 1994; Paczynski \& Xu 1994; Yu
\& Dai 2009; Yu, Wang \& Dai 2009). We give a brief description of
our model as follow. Shells with different Lorentz factors and
densities are ejected by the central black hole. Collisions of a
pair of ejecta can produce different intensities of X-rays. For
example two shells with similar Lorentz factor and density will
produce a weak flare whereas two shells with large differences can
produce a strong flare. Since the collision frequency of these pairs
(internal shocks) should be very high, it should result in rapid
variable intensities. Simultaneously some earlier ejected fast
moving shells can already reach the interstellar medium(ISM) and
produce the external shock there (Sari \& Piran 1995). Since the
earlier ejected shells are less and hence the radiation results from
external shock should be weak in the beginning. However after
collisions of pairs they can merge and move toward the ISM and
provide more energy into the external shock. Therefore the radiation
intensity due to the external shock should gradually increase. We
should note that since the injected energy provided in this way is
in a discrete manner, therefore the flux will increase substantially
but gradually decrease back to the original light curve. This
phenomenon is similar to that observed in GRBs known as
``re-brightening" effect (Zhang \& M\'{e}sz\'{a}ros 2002; Huang et
al. 2006). In Figure~\ref{LC} we provide a schematic illustration of
our model.

\subsection{Shock dynamics}
At a time of $t_{A}$, the central engine ejects a shell denoted by
$A$ with bulk Lorentz factor $\gamma_{\rm A}$ with isotropic kinetic
energy luminosity $L_{\rm A}$. Some time ($\delta t$) later, another
shell $B$ with $\gamma_{\rm B}$ and $L_{\rm B}$ is assumed to be
ejected. In order to let shell $B$ catch up and collide with the
shell $A$, $\gamma_{\rm B}>\gamma_{\rm A}$ is required. At the
radius $R_{\rm col}={\beta_{\rm A}\beta_{\rm B}c\delta
t/\psi(z)(\beta_{\rm B}-\beta_{\rm A}})$, a collision between $A$
and $B$ takes place. For $(\gamma_{\rm A},\gamma_{\rm B})\gg1$, The
collision radius is (Yu \& Dai 2009)
\begin{equation}
R_{\rm col}\simeq{2\gamma_{\rm A}^2c\delta t\over (1-(\gamma_{\rm
A}/\gamma_{\rm B})^2)\psi(z)},
\end{equation}
where $\psi(z)=1+z$.

After the collision, a forward shock and a reverse shock are
produced. The system is separated into four regions by the two
shocks and a contact discontinuity surface: (1) unshocked shell $A$,
(2) shocked shell $A$, (3) shocked shell $B$, and (4) unshocked
shell $B$, bulk Lorentz factors of which are $\gamma_1=\gamma_{\rm
A}$, $\gamma_2=\gamma_3\equiv\gamma$, and $\gamma_4=\gamma_{\rm B}$.
Two relative Lorentz factors of the shocked regions relative to
unshocked regions 1 and 4, can be calculated by
\begin{equation}
\gamma_{21}={1\over2}\left({\gamma_1\over\gamma}+{\gamma\over\gamma_1}\right),~~
\gamma_{34}={1\over2}\left({\gamma\over\gamma_4}+{\gamma_4\over\gamma}\right).\label{relgam}
\end{equation}
According to Blandford \& McKee (1976), the internal energy
densities of the two shocked regions are
$e_2=(\gamma_{21}-1)(4\gamma_{21}+3)n_1m_pc^2$ and
$e_3=(\gamma_{34}-1)(4\gamma_{34}+3)n_4m_pc^2$, where $n_1=L_{\rm
A}/4\pi R_{\rm col}^2\gamma_{\rm A}^2m_pc^3 $ and $n_4=L_{\rm
B}/4\pi R_{\rm col}^2\gamma_{\rm B}^2m_pc^3 $. The mechanical
equilibrium between the two shocked regions requires $e_2=e_3$, so
\begin{equation}
{(\gamma_{21}-1)(4\gamma_{21}+3)\over(\gamma_{34}-1)(4\gamma_{34}+3)}={n_4\over
n_1}=\left({L_4\over L_1}\right)\left({\gamma_1\over
\gamma_4}\right)^2\equiv f,\label{dyne}
\end{equation}
where $L_1=L_{\rm A}$ and $L_4=L_{\rm B}$. We can calculate the
values of $\gamma$, $\gamma_{21}$ and $\gamma_{34}$ from equations
(\ref{relgam}) and (\ref{dyne}) after the parameters of shells are
given. In four limit cases, these equations can be solved
analytically (Yu \& Dai 2009). For $\gamma_4\gg\gamma_1$, (1) if
${L_4/ L_1}\gg{(1/7)}\left({\gamma_4/\gamma_1}\right)^4$,  we have
$\gamma_{21}={\gamma_4/2\gamma_1}\gg1$,
$\gamma_{34}-1\approx{\gamma_4^2/7f\gamma_1^2}$ and
$\gamma=\gamma_4(1-\sqrt{2\xi})$, which means the forward shock is
relativistic and the reverse shock is Newtonian; (2) if $16\ll{L_4/
L_1}\ll{(1/16)}\left({\gamma_4/\gamma_1}\right)^4$, we can obtain
$\gamma_{21}={f^{1/4}\gamma_4^{1/2}/2\gamma_1^{1/2}}\gg1$,
$\gamma_{34}={\gamma_4^{1/2}/2f^{1/4}\gamma_1^{1/2}}\gg1$ and
$\gamma=f^{1/4}\gamma_1^{1/2}\gamma_4^{1/2}$, so both the two shocks
are relativistic; (3) if ${L_4/ L_1}\ll7$, we get
$\gamma_{21}-1\approx{f\gamma_4^2/7\gamma_1^2}=\xi$,
$\gamma_{34}={\gamma_4/2\gamma_1}$ and
$\gamma=\gamma_1(1+\sqrt{2\xi})$, so the forward shock is Newtonian
and the reverse shock is relativistic. Finally, (4) for
$\gamma_4\approx\gamma_1$, both the two shocks are Newtonian. Since
$\gamma_1$, $\gamma_4$, and $f$ are unchanged with the moving of the
shells, the values of $\gamma$ ,$\gamma_{21}$ and $\gamma_{34}$ are
constant before the shocks cross the shells (Yu \& Dai 2009).

\subsection{Synchrotron emission from forward and reverse shocks}
Following Dai \& Lu (2002), the total number of the electrons
swept-up by the forward and reverse shocks during a period of
$\delta t$ can be expressed by $N_{e,2}={2\sqrt{2\xi} L_A\delta
t/\left(\psi(z)\gamma_1m_pc^2\right)}$ and $N_{e,3}={L_B\delta
t/\left(\psi(z)\gamma_4m_pc^2\right)}$, respectively (Yu, Wang \&
Dai 2009).

The forward and reverse shocks can accelerate particles to high
energies. Following Sari et al. (1998), we assume that the energies
of the hot electrons and magnetic fields are fractions $\epsilon_e$
and $\epsilon_B$ of the total internal energy, respectively. Thus,
the strength of the magnetic fields is $B_i=\left(8\pi
\epsilon_{B,i}e_i\right)^{1/2}, ~~i=2,3$. We assume a power-law
distribution of the shock-accelerated electrons,
$dn_e/d\gamma_e\propto\gamma_e^{-p}$ for $\gamma_e\geq\gamma_{e,m}$
(Sari et al. 1998). The random Lorentz factor of electrons in
regions 2 or 3 is determined by
$\gamma_{e,m,i}=\epsilon_{e,i}{m_p\over
m_e}{(p-2)\over(p-1)}(\Gamma-1)$, where $\Gamma$ equals to
$\gamma_{21}$ or $\gamma_{34}$. In both shocked regions, the hot
electrons with energies above $\gamma_{e,c,i}m_ec^2$ lose most of
their energies during a cooling time $\delta t$, where the cooling
Lorentz factor is determined by ${\gamma}_{e,c,i}={6\pi
m_ec\psi(z)/\left(\sigma_T{B}_i^2\gamma \delta t\right)}$. The two
characteristic frequencies and a peak flux density are (Sari et al .
1998; Wijers \& Galama 1999)
\begin{eqnarray}
\nu_{m,i}={ q_e\over2\pi m_ec\psi(z)}{B}_i{\gamma}_{e,m,i}^2\gamma,~~\nonumber\\
\nu_{c,i}={ q_e\over2\pi
m_ec\psi(z)}{B}_i{\gamma}_{e,c,i}^2\gamma,\nonumber\\
F_{\nu,\max,i}={
3\sqrt{3}\Phi(p)\psi(z)N_{e,i}m_{e}c^2\sigma_{T}{B}_{i}\gamma\over
32\pi^2 q_e d_{L}^2}, \label{vmc}
\end{eqnarray}
where
$d_L=c(1+z)/H_0\int_0^z\frac{dz'}{\sqrt{\Omega_M(1+z')^3+\Omega_\Lambda}}$
is the luminosity distance of the source and $\Phi(p)$ is a function
of $p$, for $p=2.2$, $\Phi(p)\approx 0.6$ (Wijers \& Galama 1999).
In the calculation, we use $\Omega_M=0.3$, $\Omega_\Lambda=0.7$ and
$H_0$=70 km~s$^{-1}$~Mpc$^{-1}$. $q_e$ is the electron charge and
$\sigma_T$ is the Thomson cross section. The synchrotron spectrum
can be written as (Sari et al. 1998)
\begin{equation}
F_{\nu,i}=F_{\nu,\max,i}\times\left\{
\begin{array}{ll}
\left({\nu\over\nu_l}\right)^{1/3},~~~~~~~~~~~~~~~~~~~~~~\nu<\nu_l;\\
\left({\nu\over\nu_l}\right)^{-(q-1)/2},~~~~~~~~~~~~~~~~\nu_l<\nu<\nu_h;\\
\left({\nu_h\over\nu_l}\right)^{-(q-1)/2}\left({\nu\over\nu_h}\right)^{-p/2},~~~~\nu_h<\nu,\\
\end{array}\right.
\end{equation}
where $\nu_{l}=\min(\nu_{m,i},\nu_{c,i})$,
$\nu_{h}=\max(\nu_{m,i},\nu_{c,i})$, and $q=2$ for
$\nu_{c,i}<\nu_{m,i}$ and $q=p$  for $\nu_{c,i}>\nu_{m,i}$.

\begin{figure}
\begin{center}
\includegraphics[width=0.5\textwidth]{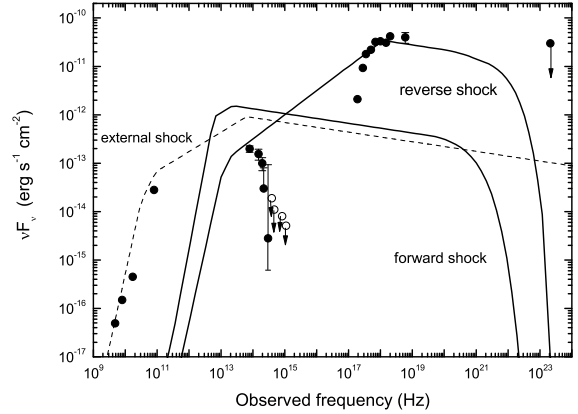} \caption{
The broadband spectral energy distribution of Swift J1644+57 at 2.9
day after the BAT trigger. The points show the observation data,
which are taken from Bloom et al. (2011). The two solid lines
represent the unabsorbed spectrum of the reverse and forward shocks,
which are generated by the internal shock. The dashed line shows the
spectrum of external shock. In order to fit the spectrum, a moderate
extinction ($A_V=3-5$) is required. } \label{spectrum}
\end{center}
\end{figure}

\section{Implication for the Swift J1644+57}
There are two peaks in the spectrum of Swift J1644+57, far-infrared
(FIR) and hard X-ray peaks. In order to fit the spectrum, we focus
on the case (3) of internal shock model in section 2.1, in which the
reverse shock is relativistic and the forward shock is Newtonian. In
the rest of the paper we denote $Q=10^xQ_x$. For illustration
purpose, we set $L_4=L_1=L=10^{47.0}\rm erg~s^{-1}$,
$\gamma_4=1000$, $\gamma_1=10$,
$\epsilon_{e,2}=\epsilon_{e,3}=\epsilon_{e}=0.5$ and
$\epsilon_{B,2}=\epsilon_{B,3}=\epsilon_{B}=0.1$. \textbf{As shown
in Cheng \& Lu (2001), the Lorentz factor of shell can be up to
$1000$. It is reasonable to adopt $\gamma_4=1000$.} According to
observation, we use $\delta t\sim 100$~s, the variability timescale
of flare (Bloom et al. 2011; Burrows et al. 2011). The collision
radius is $R_{\rm col}\sim 2\gamma_1^2c\delta t/\psi(z)\sim5\times
10^{14}$cm, which is consistent with the X-ray emission radius
determined from observation (Bloom et al. 2011). The Lorentz factor
of merged shell is $\gamma\sim 14$. Using equation (\ref{vmc}), we
can obtain the following expressions for the reverse shock
\begin{eqnarray}
\nu_{m,3} &&\simeq
1.2\times10^{18}{\rm~Hz}~\epsilon_{e,-0.3}^2\gamma_{4,3}^2L_{47}^{1/2}\epsilon_{B,-1}^{1/2}\delta
t_2^{-1}\gamma_{1,1}^{-4},\nonumber
\\  \nu_{c,3} &&\simeq
2.2\times10^{13}{\rm~Hz}~L_{47}^{-3/2}\epsilon_{B,-1}^{-3/2}\delta
t_2\gamma_{1,1}^{8},\nonumber \\ F_{\nu,\max,3} &&\simeq
0.9{\rm~mJy}~L_{47}^{3/2}\epsilon_{B,-1}^{1/2}\gamma_{1,1}^{-2}\gamma_{4,3}^{-1}d_{L,27.7}^{-2}.
\end{eqnarray}
For the forward shock, we obtain
\begin{eqnarray}
\nu_{m,2}&&\simeq
3.5\times10^{11}{\rm~Hz}~\epsilon_{e,-0.3}^2L_{47}^{1/2}\epsilon_{B,-1}^{1/2}\delta
t_2^{-1}\gamma_{1,1}^{-2}, \nonumber \\
\nu_{c,2}&&\simeq
2.2\times10^{13}{\rm~Hz}~L_{47}^{-3/2}\epsilon_{B,-1}^{-3/2}\delta
t_2\gamma_{1,1}^{8}, \nonumber \\
F_{\nu,\max,2} &&\simeq
15{\rm~mJy}~L_{47}^{3/2}\epsilon_{B,-1}^{1/2}\gamma_{1,1}^{-3}d_{L,27.7}^{-2}.
\end{eqnarray}
Therefore, the resulting synchrotron photons emitted by the two
shocks are expected to peak at two different energy bands and thus
two distinct spectral components\footnote{We can see that these
characteristic frequencies, i.e. $\nu_c$ and $\nu_m$, are very
sensitive to the Lorentz factor. However, Kobayashi et al. (1997)
have shown that the radiation loss is less than 10\% of total
energy, therefore there is virtually no evolution of these spectral
parameters during the collision.}.The peak of reverse shock spectrum
will be at hard X-ray, but peak of the forward shock will be at FIR.
The synchrotron self-absorption must be taken into account. In
$\nu_{m,2}<\nu_{a,2}<\nu_{c,2}$ case, the synchrotron
self-absorption frequency in region 2 reads (Panaitescu \& Kumar
2000)
\begin{eqnarray}
\nu_{a,2}&=&\left(\frac{5q_eN_{e,2}}{4\pi R_{\rm
col}^2B_2\gamma_{e,m,2}^5}\right)^{2/(p+4)} \nu_{m,2} \nonumber \\&&
\simeq
6.0\times10^{12}{\rm~Hz}~\epsilon_{e,-0.3}^{\frac{2p-2}{p+4}}L_{47}^{\frac{6+p}{2(p+4)}}
\epsilon_{B,-1}^{\frac{p+2}{2(p+4)}}\gamma_{1,1}^{-\frac{12+2p}{p+4}}.
\end{eqnarray}

In $\nu_{a,3}<\nu_{c,3}<\nu_{m,3}$ case, the synchrotron
self-absorption frequency in region 3 can be calculated by
(Panaitescu \& Kumar 2000)
\begin{eqnarray}
\nu_{a,3}&=&\left(\frac{5q_eN_{e,3}}{4\pi R_{\rm
col}^2B_3\gamma_{e,c,3}^5}\right)^{3/5} \nu_{c,3} \nonumber \\&&
\simeq
1.0\times10^{13}{\rm~Hz}~\epsilon_{B,-1}^{6/5}L_{47}^{8/5}\gamma_{4,3}^{-3/5}\gamma_{1,1}^{-38/5}\delta
t_2^{-2}.
\end{eqnarray}
The maximum Lorentz factor is limited by the synchrotron losses and
is given by (Cheng \& Wei 1996)
\begin{equation}
\gamma_{M,i}\simeq (3q_e/B_i\sigma_T)^{1/2}\simeq 4\times
10^7B_i^{-1/2}. \label{gammaM}
\end{equation}
Another mechanism to restrict the maximum energy of an electron is
diffusion. It turns out that maximum Lorentz factor restricted by
diffusion is much larger than that in equation (\ref{gammaM}). So
The maximal synchrotron photon energy can be estimated (Fan \& Piran
2008)
\begin{equation}
h\nu_{M,i}\simeq \frac{hq_eB_i}{2\pi m_e
c\psi(z)}\gamma_{M,i}^2\Gamma\sim \frac{30\Gamma}{1+z}~~ \rm MeV,
\end{equation}
where $h$ is the Planck constant, $\Gamma$ equals to $\gamma_{21}$
or $\gamma_{34}$.

The spectrum of internal shock model is shown in Figure
\ref{spectrum} using above parameters during a high state. The X-ray
spectrum of Swift J1644+57 can be generated in our model. A moderate
extinction ($A_V\sim 3-5$) is required to explain the spectrum. This
value of extinction is reasonable in this case, because of this
event is arising in the nucleus of host galaxy. This value is also
consistent with that determined by Bloom et al. (2011) and Burrows
et al. (2011). Because of large value of synchrotron self-absorption
frequency, the radio emission of internal shock is suppressed. From
observations, the radio emission is from larger radius comparing
with X-ray emission (Bloom et al. 2011; Burrows et al. 2011). The
interaction between the first shell ejected by central engine and
the ISM results in an external shock. The radio emission is from
large radius and can be modeled by this external shock, similar as
GRB afterglow.

The total energy release during this initial period is about $E_{\rm
iso}\sim 10^{53}$erg (Bloom et al. 2011). The ISM density is about
$n\sim 10$cm$^{-3}$. Following Sari et al. (1998) and Bloom et al.
(2011), we obtain the synchrotron frequencies and peak flux of
external shock as follow
\begin{eqnarray}
\nu_{a}&&\simeq
2.0\times10^{10}{\rm~Hz}~\epsilon_{e,-1}^{-1}\epsilon_{B,-2}^{1/5}E_{53}^{1/5}n_{1}^{3/5}, \nonumber \\
\nu_{m}&&\simeq
3.0\times10^{11}{\rm~Hz}~\epsilon_{e,-1}^{2}\epsilon_{B,-2}^{1/2}E_{53}^{1/2}t_{\rm days}^{-3/2}, \nonumber \\
\nu_{c}&&\simeq 8.0\times10^{13}{\rm~Hz}~\epsilon_{B,-2}^{-3/2}E_{53}^{-1/2}n_1^{-1}t_{\rm days}^{-1/2}, \nonumber \\
F_{\nu,\max} &&\simeq
170{\rm~mJy}~\epsilon_{B,-2}^{1/2}E_{53}n_{1}^{1/2}t_{\rm
days}^{-3/4}d_{L,27.7}^{-2}. \label{exter}
\end{eqnarray}
The expression of $F_{\nu,\max}$ is a little different from that of
Sari et al. (1998), because the observer has already observed the
edges of the jet, as discussed in Bloom et al. (2011). The spectrum
of the external shock is shown as dashed line in figure
\ref{spectrum}, which can only produce a simple power law in X-ray
region. The radio light curve is different from $t^{-5/3}$ behavior
observed the late X-ray light curve (Giannios \& Metzger 2011;
Metzger et al. 2011) because the radio light curve should be
determined by the evolution of the external shock and has nothing to
do with the accretion rate in the disk.

So our model predicates that, during the high state (flaring), the
emission of internal shock will dominate at X-ray band and a broken
power-law spectrum is shown. During the low state (no flares), there
is no internal shocks and the emission is from the external shock.
Because the typical frequencies of external shock is low at the
first few days (see equation (\ref{exter})), the spectrum at X-ray
band exhibits as a single power-law if no energy injection happens.
In both states, the radio emission is from the external shock.

\begin{figure}
\begin{center}
\includegraphics[width=0.5\textwidth]{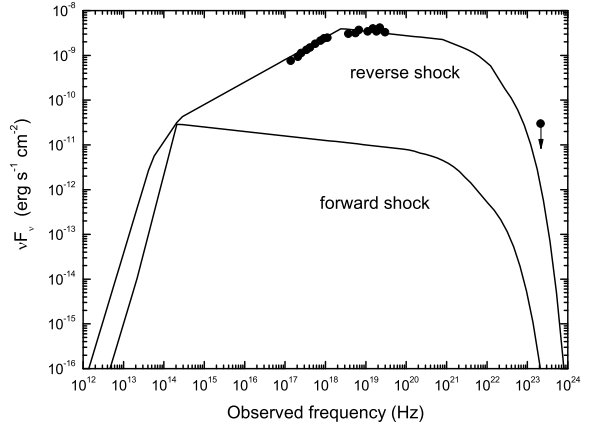} \caption{
The spectral energy distribution of Swift J1644+57 at 31 hours after
BAT trigger. The points show the observation data, which are taken
from Burrows et al. (2011). The two lines represent the unabsorbed
spectrum of the reverse and forward shocks, which are generated by
the internal shock. The parameters of two shocks are given in the
text.} \label{Burrows}
\end{center}
\end{figure}

In Figure \ref{Burrows}, we fit the spectrum data from Burrows et
al. (2011) detected at the same period. We adopt the parameters as
follow: $L_4=L_1=L=10^{48.5}\rm erg~s^{-1}$, $\gamma_4=1000$,
$\gamma_1=10$, $\delta t= 100$~s, $\epsilon_{e}=0.8$, and
$\epsilon_{B}=0.001$. Since shocks produced by collisions are highly
nonlinearly processes, therefore the microscopic parameters, i.e.
$\epsilon_e$ and $\epsilon_B$, can be different for different
collisions.

The X-ray flux from Swift J1644+57 is observed to track the X-ray
hardness (Bloom et al. 2011; Levan et al. 2011). The X-ray flux and
photon index exhibit a strong anti-correlation. This signature is a
natural consequence of our model. In the earlier stage when the the
internal shocks is dominated the X-ray flux is high but variable and
harder. At later time when the external shock is dominated, the
X-ray flux becomes lower but less fluctuating and softer.

The durations of flares are very complicated, similar as the X-ray
flares in GRBs. For individual flare, the duration can be roughly
estimated as $\Delta/c$, where $\Delta$ is the width of shell
(Maxham \& Zhang 2009). If the ejecta are coming from the disk
around the black hole, it should have the size of disk $r_d\sim
3r_s\sim 6GM/c^2\sim 8\times 10^{11}M_6~$cm, where $M$ is the mass
of black hole. From the minimum rise time, Bloom et al. (2011) and
Burrows et al. (2011) have estimated $M_6\sim10$. So the duration of
individual flare should be of order of $\Delta/c\sim
r_d/c\sim200~$s. However, flares can superimpose on each other if
shells collide near the same time. For example, the duration of the
flare detected at 111045~s after BAT trigger is about 300~s, which
is consistent with the rise time scale. But the duration of the
flare at about 1.115$\times 10^5$~s after BAT trigger with minimum
rise time is longer than 1000~s. We believe that this flare is a
superposition of several flares. It is interesting to note that the
flares detected in GRBs indicate that the shell width $\Delta$
broadens with ejected time. A natural broadening mechanism is shell
spreading. After a shell enters the spreading regime, the width of
the shell is proportional to the radius, so that if the collision
radius is larger the duration of X-ray flare can last longer. In
this event, we can also see that the width of flares broadens with
ejected time. This is similar to some central engine models of GRBs,
for example, in the fragmented disk model proposed by Perna et al.
(2006), the clumps at larger radius have lower densities and tend to
be more spread out so that the accretion time scale is longer.

\section{summary}

In this letter, we propose the internal shock model to explain the
X-ray flares of Swift J1644+57. In the internal shock model,
collisions between a series of relativistic shells generate many
pairs of forward and reverse shocks. The synchrotron emission
produced by the forward and reverse shocks could dominate at two
quite different energy bands if the Lorentz factors of these two
types of shocks are significantly different from each other. We show
that the spectral energy distribution of Swift J1644+57 could be
fitted in internal shock model, in which the reverse shock is
relativistic and the forward shock is Newtonian. A moderate
extinction ($A_V=3-5$) is required, this value is consistent with
that used in Bloom et al. (2011) and Burrows et al. (2011). Burrows
et al. (2011) showed that the high frequency spectrum is produced by
the synchrotron and SSC mechanisms, similar to poynting
flux-dominated blazar jet model. The radio fluxes come from a larger
region of the other jet. This model requires continuous in situ
re-acceleration of electrons to maintain a low energy cut-off in the
electron distribution (Aliu et al. 2011). Bloom et al. (2011)
presented two models for the spectrum: one is two-component blazar
emission model, the other is forward shock emission from jet-ISM
interaction plus EIC emission model. But on the high frequency side,
the LAT and VERITAS upper limits require that the SSC component is
suppressed by $\gamma-\gamma$ pair production. The soft photons from
the disk outflow may provide sources for the $\gamma-\gamma$
production.

The rapid rise and decline of the light curve may indicate the
internal shock origin of these flares. The external shock is very
hard to account for the X-ray flares (Burrows et al. 2005; Fan \&
Wei 2005; Zhang et al. 2006). During the high state, the emission of
internal shock will dominate at X-ray band and a broken power-law
spectrum is shown. During the low state, there is no internal shock
and the emission is from external shock. The spectrum at X-ray band
will be shown as a single power-law if no energy injection happens.
In both states, the radio emission is from the external shock.

\section*{ACKNOWLEDGMENTS}
We thank the referee for his/her detailed and very constructive
suggestions that have allowed us to improve our manuscript. We also
thank Dr. Y. W. Yu and R. Li for useful discussion. K. S. Cheng is
supported by the GRF Grants of the Government of the Hong Kong SAR
under HKU 7011/10P. F. Y. Wang is supported by the National Natural
Science Foundation of China (grant no. 11103007), Jiangsu Planned
Projects for Postdoctoral Research Funds 1002006B and China
Postdoctoral Science Foundation funded projects (20100481117 and
201104521).

\end{document}